\documentclass{emulateapj}
\usepackage{graphicx,color}
\usepackage{amssymb,amsmath}
\usepackage{appendix}
\usepackage{natbib}
\usepackage{hyperref}
\hypersetup{colorlinks=true,linkcolor=magenta,urlcolor=cyan,citecolor=blue}
\usepackage[all]{hypcap}

\providecommand{\adsurl}[1]{\href{#1}{ADS}}

\begin{document}
\shorttitle{SL analysis of RXJ0152.7-1357}
\shortauthors{Acebron et al.}

\slugcomment{Submitted to the Astrophysical Journal}

\title{RELICS: High-Resolution Constraints on the Inner Mass Distribution of the \lowercase{z}=0.83 Merging Cluster RXJ0152.7-1357 from strong lensing}
\author{Ana Acebron\altaffilmark{1}*, May Alon\altaffilmark{1}, Adi Zitrin\altaffilmark{1}, Guillaume Mahler\altaffilmark{2}, Dan Coe\altaffilmark{3}, Keren Sharon\altaffilmark{2}, Nath\'alia Cibirka\altaffilmark{1}, Maru\v{s}a Brada\v{c}\altaffilmark{4}, Michele Trenti\altaffilmark{5,6}, Keiichi Umetsu\altaffilmark{7}, Felipe Andrade-Santos\altaffilmark{8}, Roberto J. Avila\altaffilmark{3}, Larry Bradley\altaffilmark{3}, Daniela Carrasco\altaffilmark{5}, Catherine Cerny\altaffilmark{2}, Nicole G. Czakon\altaffilmark{7},William A. Dawson\altaffilmark{9}, Brenda Frye\altaffilmark{10}, Austin T. Hoag\altaffilmark{4}, Kuang-Han Huang\altaffilmark{4}, Traci L. Johnson\altaffilmark{2}, Christine Jones\altaffilmark{8}, Shotaro Kikuchihara\altaffilmark{11}, Daniel Lam\altaffilmark{12}, Rachael C. Livermore\altaffilmark{5}, Lorenzo Lovisari\altaffilmark{8}, Ramesh Mainali\altaffilmark{10}, Pascal A. Oesch\altaffilmark{13}, Sara Ogaz\altaffilmark{3}, Masami Ouchi\altaffilmark{11,14}, Matthew Past\altaffilmark{2}, Rachel Paterno-Mahler\altaffilmark{2}, Avery Peterson\altaffilmark{2}, Russell E. Ryan\altaffilmark{3}, Brett Salmon\altaffilmark{3}, Irene Sendra-Server\altaffilmark{15,16},  Daniel P. Stark\altaffilmark{10}, Victoria Strait\altaffilmark{4}, Sune Toft\altaffilmark{17} and Benedetta Vulcani\altaffilmark{5,18}}
\altaffiltext{1}{Physics Department, Ben-Gurion University of the Negev, P.O. Box 653, Be'er-Sheva 8410501, Israel\\ * anaacebronmunoz@gmail.com}
\altaffiltext{2}{Department of Astronomy, University of Michigan, 1085 South University Ave, Ann Arbor, MI 48109, USA}
\altaffiltext{3}{Space Telescope Science Institute, 3700 San Martin Drive, Baltimore, MD 21218, USA}
\altaffiltext{4}{Department of Physics, University of California, Davis, CA 95616, USA}
\altaffiltext{5}{School of Physics, University of Melbourne, VIC 3010, Australia}
\altaffiltext{6}{Australian Research Council, Centre of Excellence for All Sky Astrophysics in 3 Dimensions (ASTRO 3D)}
\altaffiltext{7}{Institute of Astronomy and Astrophysics, Academia Sinica, PO Box 23-141, Taipei 10617,Taiwan}
\altaffiltext{8}{Harvard-Smithsonian Center for Astrophysics, 60 Garden Street, Cambridge, MA 02138, USA}
\altaffiltext{9}{Lawrence Livermore National Laboratory, P.O. Box 808 L-210, Livermore, CA, 94551, USA}
\altaffiltext{10}{Department of Astronomy/Steward Observatory, University of Arizona,
933 N Cherry Ave., Tucson, AZ. 85716 USA}
\altaffiltext{11}{Institute for Cosmic Ray Research, The University of Tokyo,5-1-5 Kashiwanoha, Kashiwa, Chiba 277-8582, Japan}
\altaffiltext{12}{Leiden Observatory, Leiden University, NL-2300 RA Leiden, The Netherlands}
\altaffiltext{13}{Geneva Observatory, University of Geneva, Ch. des Maillettes 51, 1290 Versoix, Switzerland}
\altaffiltext{14}{Kavli Institute for the Physics and Mathematics of the Universe (Kavli IPMU, WPI), The University of Tokyo, Chiba 277-8582, Japan}
\altaffiltext{15}{American School of Warsaw, Warszawska 202, 05-520 Bielawa, Poland}
\altaffiltext{16}{Department of Theoretical Physics, University of Basque Country UPV/EHU, E-48080 Bilbao, Spain}
\altaffiltext{17}{Cosmic Dawn Center, Niels Bohr Institute, University of Copenhagen, Juliane Maries Vej 30, København, DK-2100, Denmark}
\altaffiltext{18}{INAF, Osservatorio Astronomico di Padova, Vicolo Osservatorio 5, I-35122 Padova, Italy}

\begin{abstract}
  \vspace{0.02cm}
Strong gravitational lensing (SL) is a powerful means to map the distribution of dark matter. In this work, we perform a SL analysis of the prominent X-ray cluster RXJ0152.7-1357 (z=0.83, also known as CL 0152.7-1357) in \textit{Hubble Space Telescope} images, taken in the framework of the Reionization Lensing Cluster Survey (RELICS). On top of a previously known $z=3.93$ galaxy multiply imaged by RXJ0152.7-1357, for which we identify an additional multiple image, guided by a light-traces-mass approach we identify seven new sets of multiply imaged background sources lensed by this cluster, spanning the redshift range [1.79-3.93]. A total of 25 multiple images are seen over a small area of $\sim 0.4$ $arcmin^2$, allowing us to put relatively highresolution constraints on the inner matter distribution. Although modestly massive, the high degree of substructure together with its very elongated shape make RXJ0152.7-1357 a very efficient lens for its size. This cluster also comprises the third-largest sample of $z\sim6-7$ candidates in the RELICS survey. Finally, we present a comparison of our resulting mass distribution and magnification estimates with those from a Lenstool model. These models are made publicly available through the MAST archive. 
\end{abstract}
\keywords{galaxies: clusters: individual (RXJ0152.7-1357, CL 0152.7-1357)-- galaxies: high-redshift -- gravitational lensing: strong}

 \section{Introduction}\label{intro}
Colliding or merging galaxy clusters are unique laboratories that can not only shed light on structure formation \citep{Peebles1989, Planck2016}, galaxy evolution \citep{Boselli2006,Deshev2017} and scaling relations \citep{Poole2007,Krause2012} of clusters during such events, but can also put important and unique constraints on the self-interaction cross-section of the elusive dark matter \citep{Clowe2006, Bradac2008, Merten2011, Dawson2012}.\\
Thanks to recent extensive observing surveys with the \textit{Hubble Space Telescope} \citep[][Coe et al. in preparation]{Lotz2017, Postman2012}, a myriad of clusters at relatively low and intermediate redshifts are analyzed in great detail.
Strong gravitational lensing (SL) is one of the most valuable tools to gain insight on the distribution of dark matter in the core of the cluster. SL has proven to provide a determination of the total mass distribution of galaxy clusters at a percent level precision \citep[e.g.][]{Richard2010, Jauzac2015, Grillo2015, Monna2017, Limousin2016, Johnson2016, Cerny2017}, in addition to allowing us to probe the early Universe, since background galaxies are magnified by the lens \citep{richard2008, Zheng2012, Coe2013, Atek2015, Livermore2017, Bouwens2017b, Hashimoto2018}. 
However, only a few cases of massive merging galaxy clusters at higher redshifts, namely close to $z\sim1$ or above, have been extensively studied \citep{DellaCeca2000, Jee2005b,Maughan2003, Coogan2018, Paterno-Mahler2018, Khullar2018}.\\
RXJ0152.7-1357 (also known as CL 0152.7-1357 and referred to as RXJ0152 hereafter), at R.A=1h52m40s, Dec=$-13^\circ 57'19''$,  constitutes one of these well-studied laboratories, yet lacking a full strong-lensing analysis until recently. \footnote{We note that through RELICS, a Lenstool model was previously made available through MAST but we present in this work the first published full SL model of the cluster.} 
\begin{figure*}
	\centering
	\includegraphics[width=0.77\linewidth]{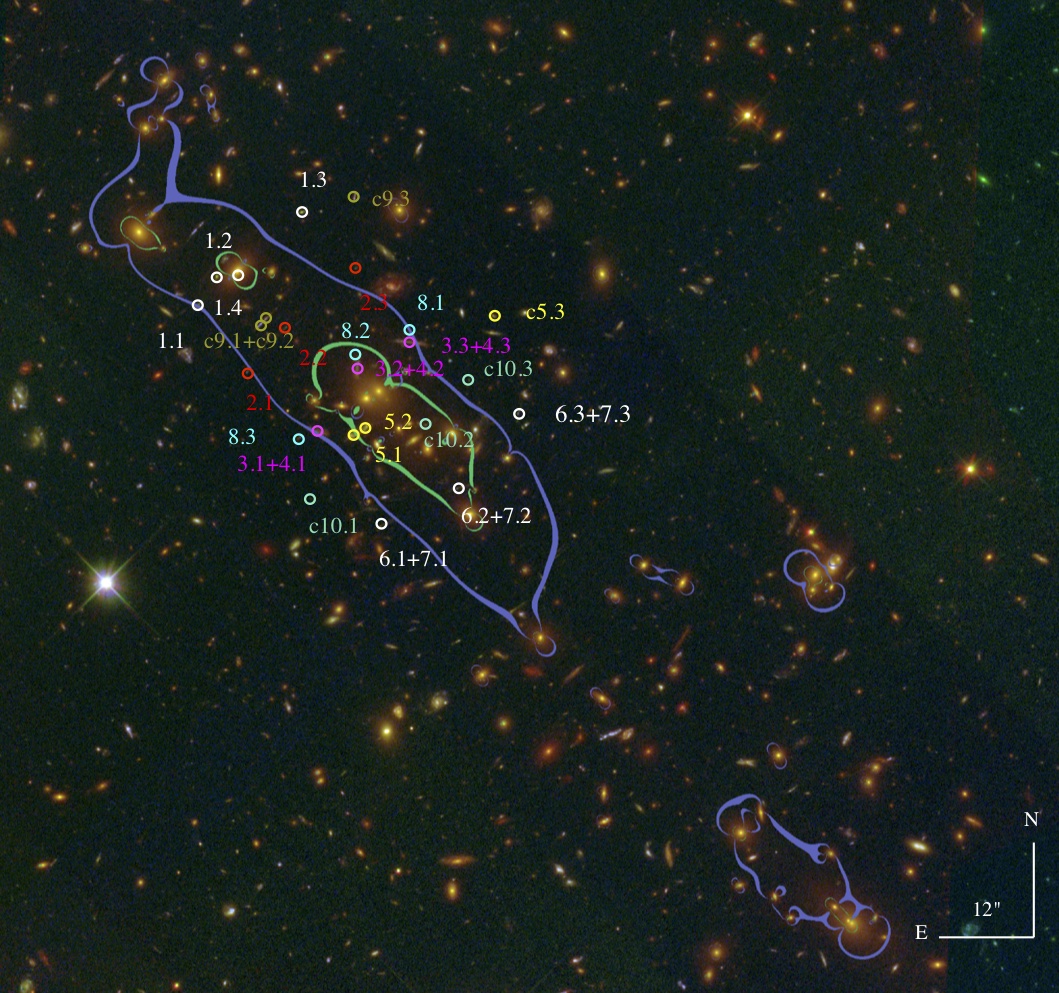} 
    \caption{Color-composite image of RXJ0152. Image was created using the \textit{HST/ACS} passbands F435W (blue), a combination of F606W+F814W (green), and a combination of the \textit{HST/WFC3IR} passbands F105W+F125W+F140W+F160W (red). The resulting critical curves from our best-fit model are displayed for a source at $z\sim 2.0$ (in green) and $z\sim 9.0$ (in violet). Multiple images (color coded to ease their identification) are labeled according to Table ~\ref{table:0308}.} 
	\label{macs0308cc}
\end{figure*} 

This cluster was detected by the \textit{ROSAT Deep Cluster Survey} \citep{Rosati1998}, the \textit{Wide Angle ROSAT Pointed Survey} \citep[WARPS,][]{Ebeling2000} and \textit{SHARC} \citep{Romer2000} as an extended source with a double core structure as well as being among the most X-ray luminous, massive merging clusters known at a redshift $> 0.55$ \citep{DellaCeca2000}. RXJ0152 was also targeted together with other 14 distant cluster candidates with the \textit{Low-Resolution Imaging Spectrograph} (LRIS) with Keck Telescope in the framework of the WARPS survey. The redshift of 6 galaxies close the X-ray peak provided a cluster redshift of z=0.8325 \citep{Ebeling2000}. Subsequent X-ray studies with \textit{BeppoSAX} \citep{DellaCeca2000} and \textit{Chandra} \citep{Maughan2003, Huo2004} found RXJ0152 to consist of two main, gravitationally bound, massive and X-ray luminous sub-clumps, at a projected distance of 730 kpc, and probably being in the early stages of a massive merging process. The X-ray temperature of the whole cluster was found to be $\sim 6.5^{+1.7}_{-1.3}$ $\mathrm{keV}$. 

RXJ0152 was also targeted through the Sunyaev-Zeldovich effect with the \textit{Berkeley-Illinois-Maryland Association} (BIMA) millimeter interferometer \citep{Joy2001}. The authors determined its total mass to be $\sim 2 \times 10^{14} \mathrm{h_{100}^{-1} M_{\odot}}$ within a $65"$ radius, consistent with the values inferred from the X-ray temperature measurements.

Extensive spectroscopic studies on this cluster that followed \citep{Demarco2005, Girardi2005, Jorgensen2005}, enabled the authors to characterize in detail the dynamical properties of this cluster, embedded into a larger-scale filamentary structure of the cosmic web \citep{Tanaka2006}. 
Dynamical studies confirmed the picture of an irregular mass distribution where cluster galaxies were observed to form substructures coinciding with those in the extended X-ray emission. These studies also indicated that the two main clumps are most likely bound and currently undergoing a merging event.

Further insights on the overall mass distribution of RXJ0152 were inferred from weak lensing (WL) studies \citep{Huo2004, Jee2005} thanks to the high-resolution of the \textit{Advanced Camera for Surveys} (ACS) observations. The WL mass estimates at a $65"$ radius were found to be in good agreement with previous results from X-ray and SZ observations. Interestingly, when comparing the WL mass reconstruction with the X-ray morphology from \textit{Chandra} and optical observations, \citet{Maughan2003} and \citet{Jee2005} noted a displacement between both the luminous and dark mass distribution and the X-ray centroids, where cluster galaxies and mass clumps seemed to lead the X-ray peaks (tracing the intracluster medium, slowed down by ram pressure). 
The fact that similar offsets are observed in other well known merging clusters \citep{Clowe2006, Markevitch2002} further strengthens the merger scenario in RXJ0152. 

In brief, previous multi-probe (X-ray, optical, SZ, dynamics and weak lensing) studies of RXJ0152 all characterized this system as highly unrelaxed and presenting a complex morphology, i.e. composed of a large number of subhaloes. 

In the central region of the cluster, where the SL features are seen, \citet{Umetsu2005} found the first multiple-image system, a $z=3.93$ galaxy lensed by the NE clump into three multiple images which allowed them to study the lensed galaxy in detail and construct a simple symmetric mass model for the NE clump \citep[with some priors drawn from previous mass estimates from WL by][]{Jee2005}. However, having only one multiple-image system usually allows one to only assess the enclosed mass within the system's effective Einstein radius, rather than to actually constrain the overall mass distribution and profile of the cluster. 

 In this work, we have taken advantage of the recent Reionization Lensing Cluster Survey (RELICS) observations \citep[e.g.][Coe et al., in prep]{Cerny2017} to revisit the SL modeling of RXJ0152, the third highest-redshift cluster of the sample. We identify various additional sets of multiple images spread throughout the central clumps, allowing us to constrain in detail the inner mass distribution of the cluster using two well-known modeling tools.

A major goal of the RELICS survey is to detect a large sample of high-redshift \citep{Salmon2017}, magnified galaxies. SL models for high-redshift clusters are of great interest as the cluster lensing power increases significantly with source redshift compared to lower-redshift counterparts \citep{Zitrin2013}. Indeed, RXJ0152 presents the third-largest sample of high-z ($z\gtrsim6$) candidates from the RELICS cluster sample \citep{Salmon2017} making the SL models presented here crucial for accurately determining the properties of these high-redshift candidates as well as translating the sample of candidates into a galaxy luminosity function.

This work is organized as follows: in \S 2 we briefly describe the observations. These were used to identify multiple images considered for the SL analysis, presented in \S 3. The results are presented and discussed in \S 4. Finally the work is summarized in \S 5. Throughout we assume a $\Lambda$CDM cosmology with $\Omega_{\rm m0}=0.3$, $\Omega_{\Lambda 0}=0.7$, $H_{0}=100$ $h$ km s$^{-1}$Mpc$^{-1}$, with $h=0.7$, where $1\arcsec= 7.71$ kpc at the redshift of RXJ0152.

\section{Data and observations} \label{sec:data}
\subsection{Imaging}
The cluster analyzed in the present work is part of the RELICS cluster sample (PI: D. Coe, Coe et al., in preparation). The RELICS program has targeted 41, mainly SZ-selected massive clusters \citep[including other several criteria, see][or Coe et al., in preparation, for more details]{Cerny2017} to efficiently search for magnified high-redshift galaxies in time for spectroscopic follow-up with \textit{James Webb Space Telescope} \citep{Salmon2017}. 
Given that some Hubble Space Telescope (\textit{HST}) archival observations already existed for RXJ0152 (program 9290 and follow-up observations searching for supernovae programs 10493 and 10793), RELICS completed the observations needed to make this cluster a coherent part of the RELICS sample. In total, including the previous existing observations, RXJ0152 has been observed for a total of 3 orbits with the \textit{Advanced Camera Survey} (ACS- in the F435W, F625W, F775W, F850LP bands), 2 orbits with  the \textit{Wide Field Camera 3} (WFC3/IR- in the F105W, F125W, F140W, F160W bands) and 30 hours per band of each of the \textit{Spitzer}-IRAC channels (PI: M. Bradac, PI: Soifer). 
In this work we used the reduced \textit{HST} images, and photometric source catalogs generated with SExtractor \citep{Bertin1996} in dual-image mode from the final drizzled 0.06" images. Bayesian photometric redshifts (hereafter $\mathrm{z_{phot}}$) were derived using the \textit{Bayesian Photometric Redshift} program \citep[BPZ,][]{Benitez2000, Benitez2004, Coe2006} from seven \textit{HST} band imaging-data (both from RELICS observations and \textit{HST} archival data). 
These data products are available for the community through the Mikulski Archive for Space Telescopes (MAST)\footnote{\url{https://archive.stsci.edu/prepds/relics/}\label{mast}}.
\\
\begin{turnpage}
\begin{table*}
	\caption{Multiple images and candidates for  RXJ0152.7-1357.}       
	\label{table:0308}      
	\centering                          
	\begin{tabular}{c c c c c c c c c}        
		\hline\hline                 
		Arc ID & R.A. & Decl& $\mathrm{z_{phot}}$ [$\mathrm{z_{min}}$-$\mathrm{z_{max}}$]\tablenotemark{a} & $\mathrm{z_{spec}}$ & $\mathrm{z_{model}^{LTM}}$[$68\%$ C.I.]\tablenotemark{b}& $\mathrm{z_{model}^{Lenstool}}$[$68\%$ C.I.]\tablenotemark{c} & Comments & individual RMS (")\tablenotemark{d}\\  
		&[deg]&[deg]&&  && \\
		\hline          
		1.1   & 28.189012 & -13.952162  & 3.79 [3.70-3.89] &3.93\tablenotemark{e} &3.93\tablenotemark{f} & 3.93 &&1.63 \\  
		1.2   & 28.188295 & -13.951124 & 3.77 [3.61-3.91]&"& "&"& &1.27 \\  
		1.3   & 28.185218 & -13.948862  &3.85 [3.77-3.94]&"& "&"&&1.35 \\
        1.4   & 28.187603 & -13.951152  & -&"&"  &"&close to a cluster member &0.38 \\  
		\hline
		2.1   & 28.187214 & -13.954498 & 2.81 [2.42-3.14] &-& 3.80 [3.03-4.23]& 3.58 [2.46-3.67] &&0.87 \\  
		2.2   & 28.185771 & -13.952901 &3.25 [2.96-3.52] &-&" &"&&0.26 \\
        2.3   & 28.183267 & -13.950841 &0.66 [0.22-3.13] &- &" &"&&0.85 \\  
		\hline
		3.1   & 28.184847 &-13.956517 &1.80 [1.54-1.98]&-& 1.98 [1.67-2.01]& 1.58 [1.50-1.68] &&0.38 \\  
		3.2   & 28.183187 & -13.954174  &1.06 [1.00-1.18]  &- & "&"&&0.96 \\  
        3.3   &28.181607  &-13.953402   &0.27 [0.03-0.46] & -&" &"&& 0.14\\  
        \hline
		4.1   &28.184489 &  -13.956624 &1.79 [1.61-1.89] &-& 1.97 [1.68-2.01]&  1.58 [1.50-1.70] & & 0.20\\  
		4.2   & 28.183217 &  -13.954531 &- &- &" &" & &1.15 \\ 
        4.3   & 28.181020 &  -13.95347 &1.79 [1.55-1.92] & -&"& "&&0.68\\
         \hline
		5.1   & 28.183455 & -13.956739  &3.27 [3.16-3.35]&-&1.79 [1.60-10.83]& 2.26 [2.02-2.63] & & 0.36\\  
		5.2   & 28.182859 & -13.956402  &0.21 [0.16-0.50] &- &" &"&& 0.85\\ 
		c5.3   & 28.178268 & -13.952506 & 3.13 [2.94-3.32] &- &" &"&& -\\ 
        \hline
		6.1   & 28.182379&  -13.959926 & 3.01 [0.09-3.26] &-&3.00 [2.39-3.10]& 2.06 [1.83-2.43] & &0.55 \\  
		6.2   &28.179398  &-13.958503   & 3.11 [2.49-3.38] &- & "&  "&&0.46 \\ 
        6.3   & 28.177391 & -13.955948 & 2.79 [2.51-3.06]& -&"& "&&0.68\\
        \hline
		7.1   &28.182153 & -13.959811  & 1.15 [1.14-3.03] &-&3.13 [2.36-3.12]& 2.02 [1.77-2.36] &&1.19 \\  
		7.2   & 28.179577 & -13.958584  & 1.16 [1.12-3.03]&- &" &" & &0.41 \\ 
        7.3   & 28.177323& -13.955818  & 2.78 [2.33-3.09] &- &"&"& &1.01\\
        \hline
		8.1   &28.181273 & -13.953073  &  2.88 [2.64-3.28]&-&2.17 [1.84-2.18]& 1.79 [1.69-1.94] &&0.49 \\  
		8.2   & 28.183334 & -13.954073  & -&-&" & "& &0.85 \\ 
        8.3   & 28.185257 & -13.956773  &1.91 [1.48-2.38] &- &"& "&&0.18\\
        \hline
		c9.1   & 28.186812&  -13.952893 &  1.91[1.82-2.37]&-& $\sim2.8$ &-&not used as constraint &- \\  
		c9.2   & 28.186540 & -13.952623 & 2.51[2.32-2.63]&-&" &-&  &- \\ 
        c9.3   & 28.183312 & -13.948390 &1.70[1.59-1.79] &- &"&-& &-\\
        \hline
		c10.1   & 28.184906&  -13.958930 &   3.82[0.19-4.12]&-&$\sim2.6$ &3.25 [2.91-3.67]& not used as constraint &- \\  
		c10.2   & 28.180724 & -13.956335  & -&-&" &-& &- \\ 
        c10.3   & 28.179147 & -13.954763 &3.40[2.92-3.67]  &- &"&-& &-\\
		\hline\hline 
	\end{tabular}
    \tablecomments{}
  \tablenotetext{1}{Photometric redshift with upper and lower limits, based on the BPZ estimates from RELICS catalog with the $95\%$ confidence range. ~~~In this column a "-" sign indicates an image for which its $\mathrm{z_{phot}}$ could not be measured due to light contamination or poor signal-to-noise ratio.}
  \tablenotetext{2}{Redshift prediction based on our LTM best-fit model.}
  \tablenotetext{3}{Redshift prediction based on our Lenstool best-fit model.}
 \tablenotetext{4}{RMS between the observed and model-predicted multiple images from our LTM best-fit model.}
 \tablenotetext{5}{\citet{Umetsu2005}}
  \tablenotetext{6}{fixed redshift for the LTM modeling.}
\end{table*}
\end{turnpage}

\subsection{Spectroscopic observations}
The cluster was observed with LDSS3-C\footnote{\url{http://www.lco.cl/telescopes-information/magellan/ instruments/ldss-3}} on the Magellan / Clay telescope on 2017 July 27 (University of Michigan allocation, PI: Sharon). The seeing ranged between $0\farcs5-0\farcs7$ with thin clouds throughout the night. The data were obtained with the VPH-ALL grism ($4250\AA < \lambda < 10000 \AA$). The spectra were reduced using the standard COSMOS routines \citep{Dressler2011,Oemler2017}. A full description of spectroscopic follow-up will be presented in a forthcoming paper (Mainali et al., in prep). 
We measure two secure redshifts in this field, both from detection of Ly-$\alpha$. An image of system 1 at 1:52:45.358, -13:57:07.75, confirming the redshift previously measured by \citet{Umetsu2005}, $z_{spec} = 3.930$, and a galaxy at 1:52:39.566, -13:58:37.11, $z_{spec}= 3.611$. 

\section{Lens model} \label{sec:lens_model}
\subsection{The LTM pipeline}\label{sec:lens_modelLTM}
We perform the SL analysis using the LTM method by \citet[][]{zitrin2009, Broadhurst2005b}. LTM has proven to be a powerful method to both identify new multiple images, and constrain the cluster mass distribution \citep[e.g.,][]{Merten2011, zitrin2015, Frye2018}. The LTM pipeline has been adopted as well to model other RELICS clusters \citep[see][]{Acebron2018, Cibirka2018}. We give here a brief overview of the pipeline, but we refer the reader to these recent papers for further details. 

Our method relies on the assumption that the underlying dark matter (DM) distribution in the cluster is traced by the distribution of the luminous component -- or namely -- cluster galaxies. This brings to a minimum the number of free parameters needed to generate a mass model, while still possessing sufficient flexibility to describe the underlying mass distribution. The position and source redshift (where available) of multiple images are used as constraints for the SL modeling. \\
We start constructing a mass model by identifying cluster members, following the red-sequence method \citep{Gladders2000}. 
We use the magnitudes measured from the F606W and F814W filters to draw a color-magnitude diagram and consider only galaxies down to 24 AB within $\pm0.3$ mag of the sequence. We then apply several criteria to exclude stars from our selection: we consider objects with magnitudes fainter than 17 AB with a cut-off value for the \textit{stellarity} index of $<0.95$ and rely as well on the help of a size-magnitude relation, plotting the FWHM versus the F814W magnitude in which stars occupy a specific region of the parameter space.
An important step is a subsequent visual inspection of the selected cluster members where we discard further interloping galaxies (bright foreground galaxies for instance) or artifacts (such as faint and diffuse objects). 
We also used the delivered photometric catalog which includes photometric redshift estimates from BPZ to check that all selected cluster members were within $z_{phot} \pm 0.1$ of the mean redshift of the cluster.
Finally, we also compared our selection with previous publicly available spectroscopic catalogs from \citet{Demarco2005,Demarco2010}. In the most central regions of the cluster, apart from the brightest central cluster members, a large number of galaxies that appear to be red-sequence cluster members were lacking a spectroscopic confirmation so we chose to rely on the red-sequence method for the rest of the analysis.\\
Once a final list of cluster members is constructed, each cluster member is then parametrized by a symmetric power-law surface mass-density distribution, scaling linearly in amplitude with luminosity (for some galaxies, ellipticity or other scaling relations can be introduced; see below).
The power-law exponent is the first free parameter of the model and the same for all galaxies.
The dark matter distribution, as is assumed in our method, will follow the luminous component as well but is smoothed with a 2D Gaussian- whose width is the second free parameter of the model. Both components are then co-added with their relative weight being the third free parameter. The fourth free parameter refers to the overall normalization. Our method allows for further flexibility by adding a two-parameter external shear (which introduces ellipticity to the magnification map) parametrized by its amplitude and its position angle, bringing to a total of 6 basic free parameters. Finally, to better reproduce the observations, other parameters can be introduced, such as the weight of the BCG, its ellipticity, position angle or redshift of background sources, which can be optimized by the pipeline.

The goodness of the fit is assessed using a $\chi^2$ criterion during the minimization which quantifies the quality of reproduction of multiple-image positions in the image plane, given by: 
	\begin{equation}
	\centering
	\chi^2 = \sum\limits_{i=1}^{n} \dfrac{( x_i^{pred} - x_i^{obs})^2 + ( y_i^{pred} - y_i^{obs})^2}{\sigma_{i}^2} \mathrm{,}
	\end{equation} 
where the the difference between the model predicted $\mathrm{x_i^{pred}}$, $\mathrm{y_i^{pred}}$ and observed positions $\mathrm{x_i^{obs}}$, $\mathrm{y_i^{obs}}$ of the multiple images is weighted by the observational uncertainty $\sigma_{i}$ (assumed here to be of $0.5"$ for all multiple images). \\
Independently, we can also assess the goodness-of-fit of a model with the root-mean-square (RMS) between the observed and model-predicted positions of the multiple images in the image plane, which can written as follows: 
\begin{equation}
RMS = \sqrt[]{\dfrac{1}{N_{img}} \sum_{i=1}^{n} \left(( x_i^{pred} - x_i^{obs})^2 + ( y_i^{pred} - y_i^{obs})^2 \right)} \mathrm{,}
\end{equation}
where $\mathrm{N_{img}}$ is the total number of images.\\

\subsection{Identification of multiple images}
In an iterative way and starting with a simple initial model, our method predicts both the shape and orientation of multiply-imaged candidates by sending them to the source plane and back to the image plane using the lens equation (namely $\beta=\theta-\alpha$, where $\beta$ is the angular source position, $\theta$ the observed image position, and $\alpha$ the so-called reduced (i.e., scaled) deflection angle, in this case given by the initial model). Using these predictions we thus can identify new multiple-image families based on similar colors, morphology and symmetry, allowing us to refine the initial model. In this study we only use as constraints the position of multiply imaged systems that we consider secure (i.e. those whose agreement with the model prediction, internal details, similar colors, and symmetry, leave essentially no doubt these are images of the same source).

\begin{figure}
	\centering
	\includegraphics[width=0.9\columnwidth]{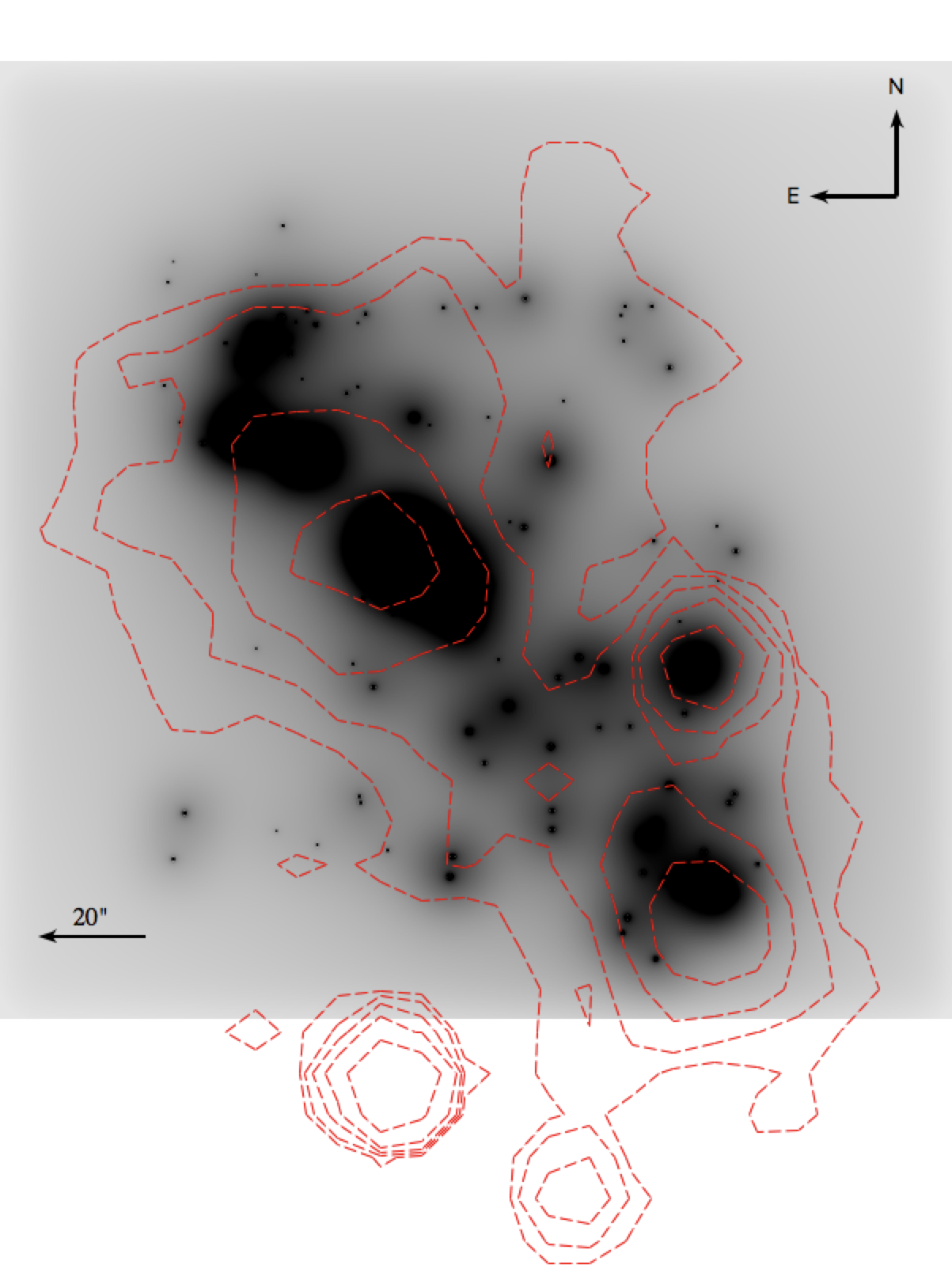}
	\caption{Left panel - Convergence $\kappa$ map from our best-fit LTM model (referring to the projected surface mass density in units of the critical density for lensing $\Sigma_{crit}$), scaled to a source redshift of $z_s\sim2.0$ where are overlaid the smoothed X-ray (red dashed contours) from previous \textit{Chandra} observations.}
	\label{kappa}
\end{figure} 
The first multiply-imaged system used to constrain the mass model was reported by \citet{Umetsu2005} who measured a spectroscopic redshift of 3.93 with the \textit{Faint Object Camera And Spectrograph} on the \textit{Subaru} telescope (FOCAS). This background galaxy is lensed by the NE clump into three multiple images appearing as a greenish galaxy on the \textit{HST} composite image in Figure~\ref{macs0308cc}. We report however an additional image in the system, image 1.4, which lies next to a nearby cluster member. This is the only system having a redshift spectroscopically confirmed. 
In addition we have identified seven other multiple-imaged systems, displayed in Figure~\ref{macs0308cc} and listed in Table~\ref{table:0308}.
System 2 comprises three multiple images that have similar colors in the \textit{HST} composite image. Systems 3 and 4, with three multiples images each, show an arc shaped image in the central region of the cluster. Due to the difference in color between the two ends of the arc we mark them as two different systems, supplying two sets of constraint to the model. System 5 consists of three multiple images, images 5.1 and 5.2 are stretched into an arc shape, with two, bright emission knots, appearing light green in a composite color image as in Figure~\ref{macs0308cc}. A candidate counter-image c5.3 sits on the other side of the opposite critical curve. As seen in Table~\ref{table:0308}, the redshift of this system is not well constrained with the LTM optimization. 
We use a predicting tool to de-lens one image of the system to the source plane and back to the image plane to compare the model-predicted and observed location and orientation reproduction. Our best-fit LTM model for prefers a higher redhsift (in the redshift range $2.5-3.1$), in good agreement with the Lenstool and BPZ estimations.

\begin{figure*}
	\centering
	\includegraphics[width=.7\linewidth]{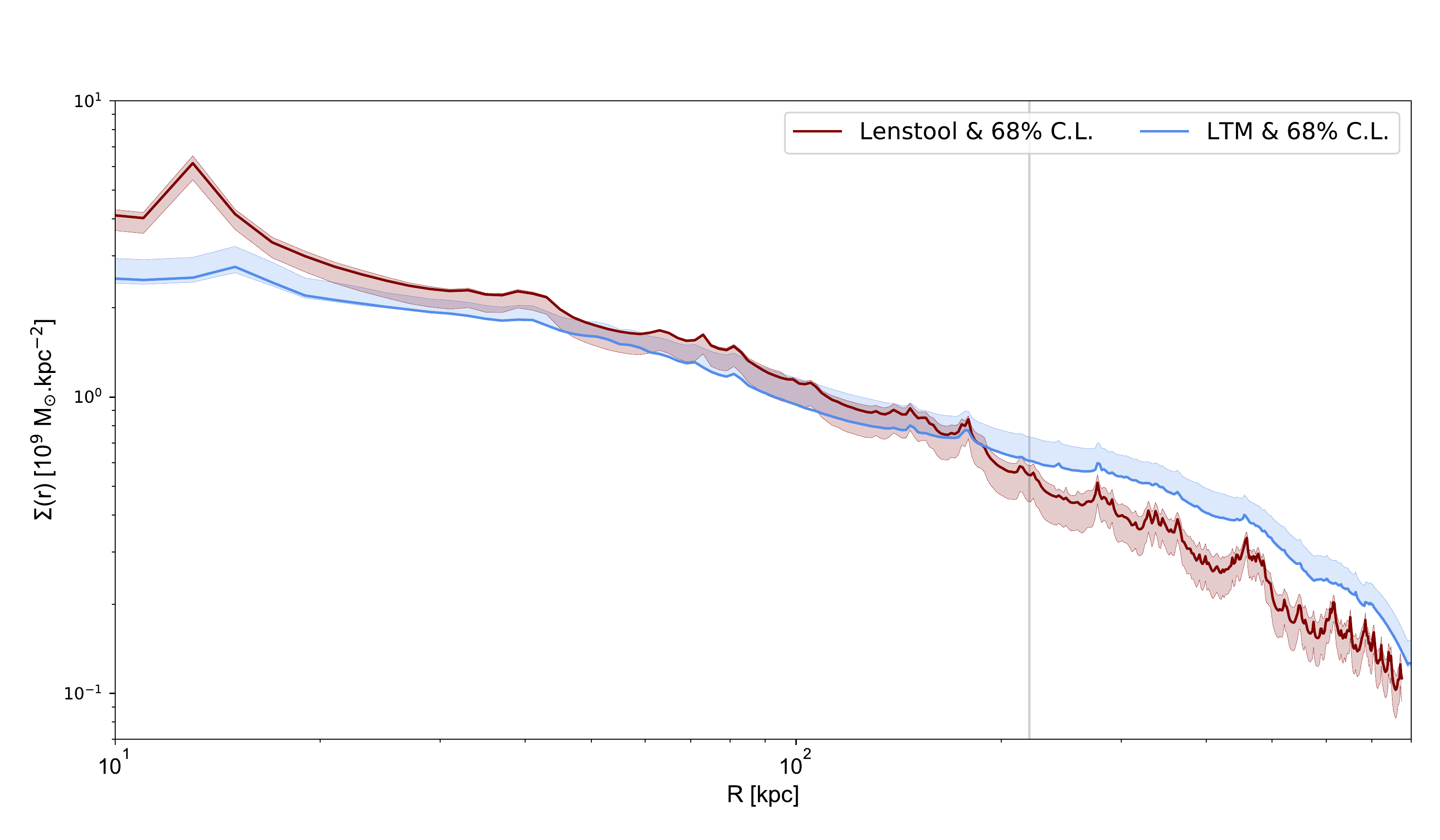} 
	\caption{The mass profile computed as the mass within an annulus at a certain radius for the LTM and Lenstool models in blue and red, respectively. Both profiles are centered in one of the BCGs. The black dashed vertical line sets the radius within which we have identified multiple images.}
	\label{mass}
\end{figure*} 

The three multiple images making up system 6 and system 7 appear as two bright peaks with similar colors, respectively, lying next to each other. Finally, systems 8 has three multiple images that appear as a bright peak with a long tail. All multiple images are marked in Figure~\ref{macs0308cc} and, their reproduction by our best-fit model, is shown in Figure~\ref{stamps0308}. Our model only predicts two additional, fainter, multiple images for system 5. 
Other, less secure, multiply imaged systems predicted by our SL model are reported in Table~\ref{table:0308} as candidates. System c9 appears as 3 images, one of them being a pink arc, with several emission knots (images c9.1 and c9.2). System c10 comprises 3 green images and is considered as a candidate as few nearby objects are similar in terms of colors and morphology, diminishing the reliability of our candidate identification. Therefore, we chose not to include these counter images in the modeling and only refer to them as possible candidates. \\
We find that the best SL model for RXJ0152 is obtained when not considering any galaxy as predominant (i.e a BCG) as its structure shows a very elongated cluster with no clear central, predominant region. Typically, with the LTM formalism, the BCG is found to contain more mass with respect to its light compared to other members, and we therefore usually allow its M/L ratio to vary while we found this was not needed for the modeling of RXJ0152. We also do not assign any ellipticity for the central bright galaxy. We do however optimize both the ellipticity and position angle of the bright cluster member of the NE clump, close to the images of system 1, which slightly improves their reproduction.

We scale our model to the spectroscopic redshift of system 1 (see Table~\ref{table:0308}) and leave the redshift of the remaining systems as free parameters to be optimized in the minimization procedure (allowing the corresponding $\mathrm{D_{LS}}/\mathrm{D_S}$ ratio for each system to vary by $-0.3$ and $+0.5$). \\
The optimization of the model is carried out with several
thousand Markov Chain Monte Carlo (MCMC) steps and includes a total of 14 free parameters when accounting, in addition, for freely optimized galaxies and source redshifts where needed.

The resulting critical curves (for a source at $z_s = 2$ and $z_s = 9$) for our final best-fit model, which has an image reproduction $RMS =0.84\arcsec$, are shown in Figure~\ref{macs0308cc}. The reproduction of the multiple images used as constraints in our model are shown in Figure~\ref{stamps0308} and the obtained best-fit parameters are presented in Table \ref{table:BFltm} which are specific to our methodology as the LTM model is not analytic.

\subsection{The Lenstool pipeline}

RXJ0152.7-1357 was also modeled with the Lenstool\footnote{\url{https://projets.lam.fr/projects/lenstool}} pipeline \citep[see][for further details]{Jullo2007} and model products were made publicly available by the RELICS team through MAST. In order to compare the main SL outputs between the two modeling algorithms, we revisit the Lenstool analysis and compute a second version (V.2) using the same lensing constraints as the LTM model, except for system 9 that is in addition included in the Lenstool model. We provide here a comparison of the main SL outputs between the LTM and Lenstool pipelines but refer the reader to a forthcoming study for a more detailed and extensive comparison \citep[see also][for comparison studies of different SL algorithms]{Meneghetti2017, Remolina2018}.

RXJ0152 is modeled using the same constraints reported in Table ~\ref{table:0308} except for system 9 that is included in the Lenstool modeling (images 9.1 and 9.2). Both the large and small-scale haloes are parametrized by a pseudo isothermal density profile \citep[PIEMD,][]{Kassiola1993}. We optimize the ellipticity, position angle, core radius and velocity dispersion of the main large-scale halo; the central coordinates of the halo are also let free during the optimization.
Moreover, the cluster member close to system 1 is modeled independently with a PIEMD profile where the core radius and velocity dispersion are optimized during the minimization procedure (i.e. not following the scaling relations). 
The LTM pipeline provides a hint on a additional mass in the outskirts of the cluster in the South-West direction that can also be seen in the X-ray map (see Figure~\ref{kappa}). We found that an additional large-scale clump for this SW structure improved the Lenstool fit by $\sim0.1"$. However, this improvement is not significant enough, in terms of the \textit{Bayesian Information Criterion} (B.I.C.) that prefers a model not including additional free parameters to be optimized in a region with no lensing constraints. 

Finally, the small scale haloes associated with cluster members are also parametrized with a PIEMD profile with a fixed core radius of 0.01 kpc, a velocity dispersion allowed to vary between 50 and 200 km/s, a cut radius varying from 20 kpc to 200 kpc and following the scaling relations \citep{Faber1976}. We assign the ellipticity and position angle values measured from the light distribution with SEXtractor \cite{Bertin1996} to model the underlying dark matter distribution. As in the case of our modeling with LTM, the redshift of all systems but system 1 are optimized with a flat prior.

Our resulting best-fit model from Lenstool has an RMS of $0.52\arcsec$. The best fit parameters are shown in Table \ref{table:BF2} and the resulting critical curves and magnification map from our best-fit model are shown in Figure ~\ref{cc_LT}.
We have also carried out a model where the galaxy-haloes are considered spherical (as in the LTM model). The resulting RMS is of $0.59\arcsec$, which is very similar to our fiducial model and their mass profiles are also equivalent within the statistical uncertainties. 
Finally, we have also modeled RXJ0152 without system 9 with Lenstool. This model yields an equivalent fit (with an RMS of $0.53\arcsec$) in terms of best fit parameters to that of our fiducial model but the latter yields a more robust mass profile estimation in the inner cluster region, i.e. with lower statistical uncertainties.

\section{Results and discussion} \label{sec:results}
Both the surface mass-density distribution from our best-fit model and the mass profile are shown in Figure ~\ref{kappa} and ~\ref{mass}, respectively. Our SL analysis reveals, as implied by its member galaxy distribution, a highly elongated cluster in the NE-SW direction (see the $\kappa$ map in the left panel of Figure ~\ref{kappa}), composed of several clumps. 
We also compare the mass distribution of RXJ0152 obtained from our SL analysis to previous high-resolution X-ray observations with \textit{Chandra} \citep[ObsId 913,][]{Ebeling2000, Maughan2003} in Figure ~\ref{kappa} (see dashed red contours).

While our LTM pipeline strongly follows the assumption that light traces mass, \citet{Jee2005} showed in their WL analysis that there exists a strong correlation between both the light and mass components. However, the X-ray peaks are displace with respect to the peaks of the mass distribution \citep[see also][]{Maughan2003}.
Together with previous X-ray \citep{Maughan2003} and WL studies \citep{Jee2005}, the elongated, filamentary-like structure of the SL region, further supports the merging scenario. 

We compute the effective Einstein radius of RXJ0152, defined as $\theta_E=\sqrt[]{A/\pi}$, where $A$ refers to the area enclosed within the critical curves. Our SL analysis reveals a relatively small lens, with an effective Einstein radius of $\theta_E(z_s=2)= 8.5\pm 1\arcsec$ and a corresponding enclosed mass within the critical curves of $ 2.5\pm0.4 \times 10^{13} M_{\odot}$ \citep[with the uncertainties typically encompassing both the statistical and systematic errors, e.g.][]{zitrin2015}. The critical area increases significantly for higher redshift sources due to the merging of the critical curves by different clumps as shown in Figure ~\ref{macs0308cc}, reaching $\theta_E(z_s=9)= 19\pm 2\arcsec$. We also find an Einstein radius of $\theta_E(z_s=3.93)= 15.2\pm 1.5\arcsec$, in agreement with the previous estimation from \citet{Umetsu2005}.

\begin{figure*}
	\centering
	\includegraphics[width=0.50\linewidth]{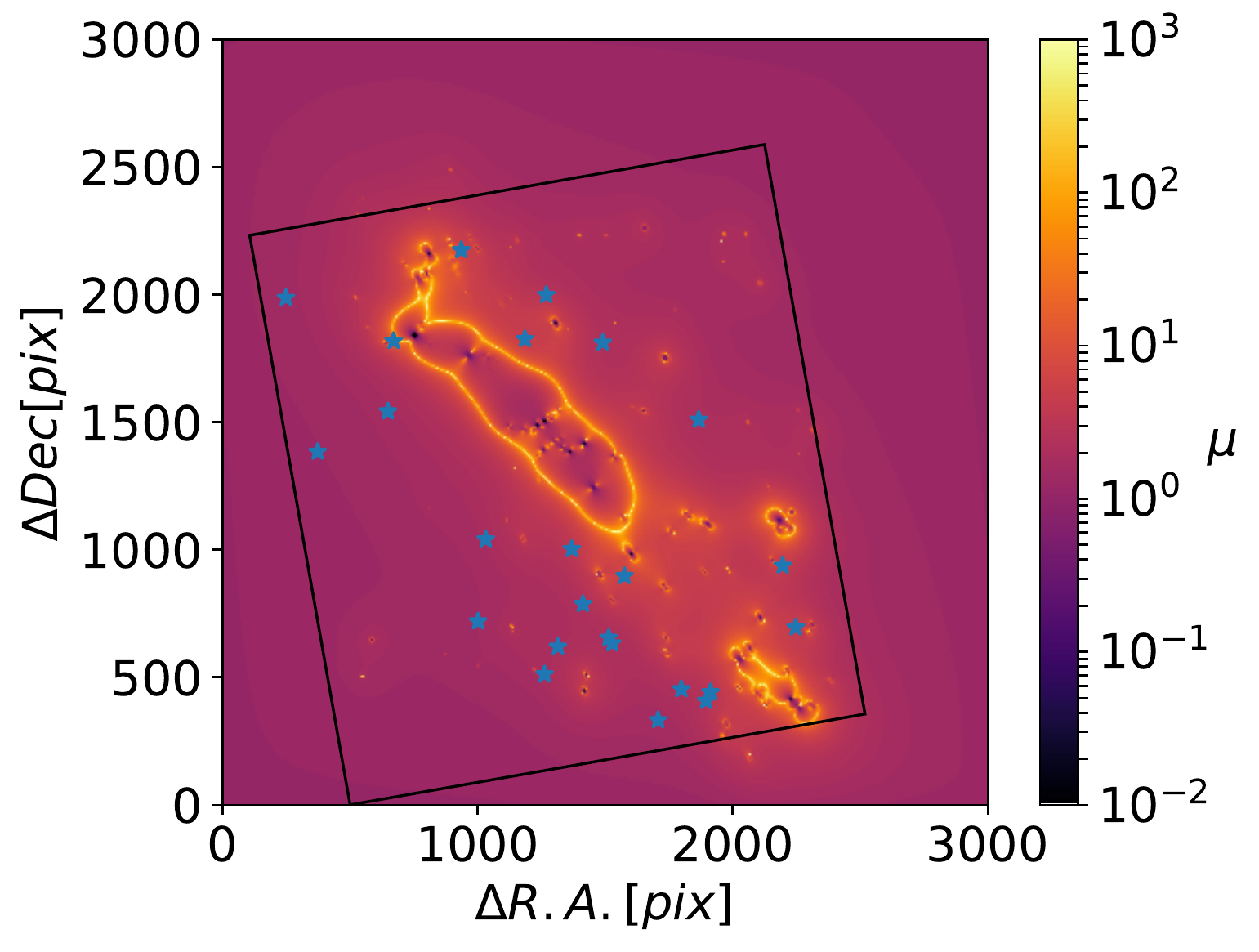} \hfill
	\includegraphics[width=0.49\linewidth]{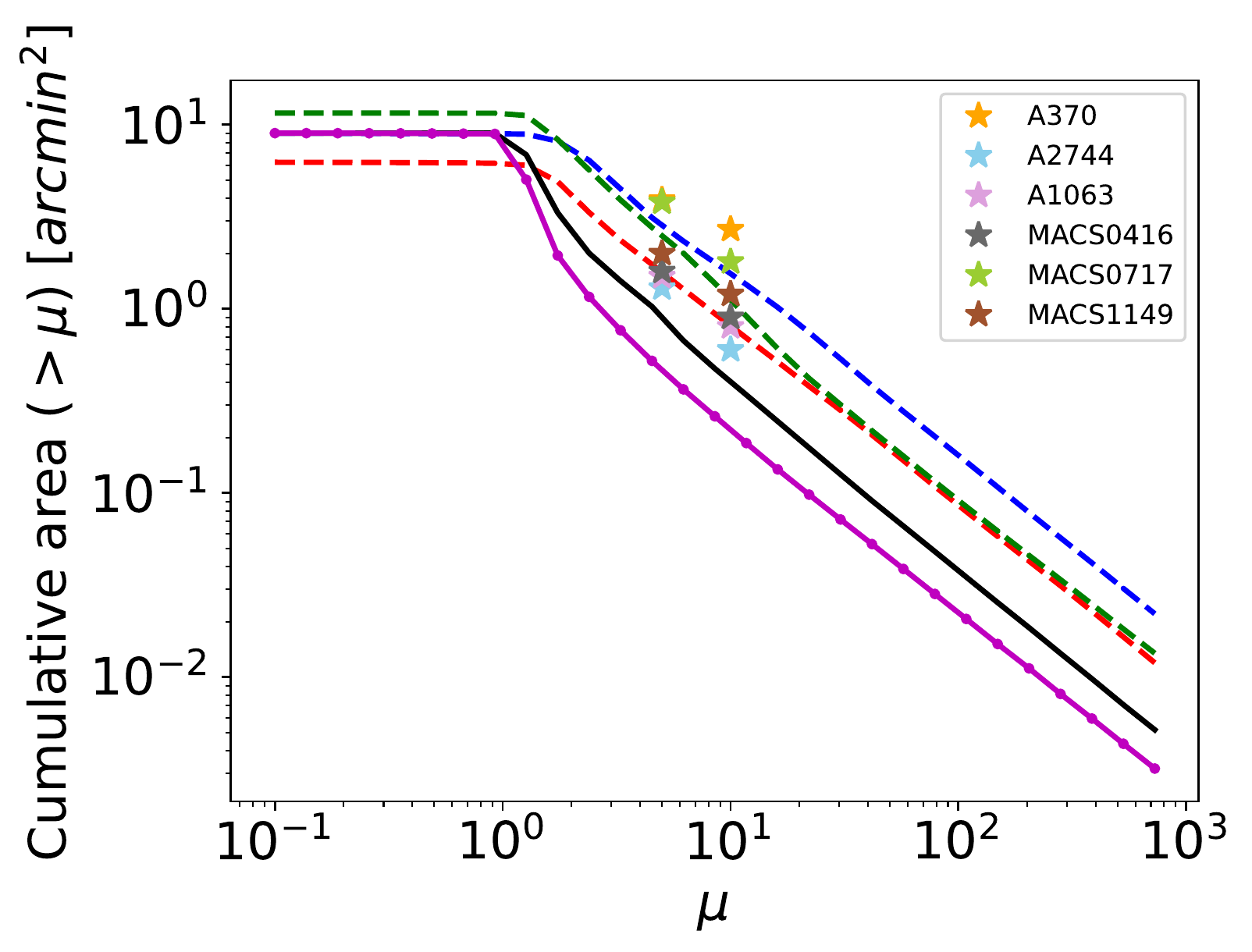} 
	\caption{Left panel: Magnification map from our best-fit LTM model for a source at $z_s=6.5$, the redshift around which the majority of the RELICS high-z candidates were found in \citet{Salmon2017}- pictured as blue stars. The black rectangle indicates the WFC3/IR FOV with a different roll angle to fit in all high-z candidates. Right panel: we also assess the strength of RXJ0152 as a lens, comparing the cumulative area having a magnification higher than a given value for a source at $z_s=9.0$ (in black and magenta for our LTM and Lenstool models, respectively) with those from other known efficient RELICS SL clusters modeled with the LTM pipeline, MACS J0308.9+2645, PLCK G171.9-40.7 and Abell S295 (in blue, red and green, respectively). The cumulative areas ($\mu>5$ and $\mu>10$) for the \textit{Hubble Frontier Fields} clusters are also indicated as colored stars, computed from the submitted \textsc{zitrin-ltm-gauss} models. The $1\sigma$ errors are typically of the size of the star symbol.}
	\label{mag}
\end{figure*}

Figure ~\ref{mag} shows the magnification map for a source at redshift $z_s=6.5$ (left panel) together with the position of the high-z candidates reported by \citet{Salmon2017} within RXJ0152's field of view. Overall we find that our magnification estimation is constrained to better than $20\%$ in at least $80\%$ of the modeled FOV. The cumulative area magnified above a certain magnification value (which assess the strength of the lens) at a source redshift $z_s=9.0$ is shown in the right panel. RXJ0152 covers a modest area of high-magnification, $\sim 1.05$ arcmin$^{2}$, for $\mu > 5$ to $\sim 0.47$ arcmin$^{2}$ for $\mu=10$, for a source at a redshift of $z_s=9.0$. 
The cumulative area measured for RXJ0152 is compared to other RELICS clusters that provided a large high-magnification area, MACS J0308.9+2645 and PLCK G171.9-40.7, presented in \citet{Acebron2018} and Abell S295 \citep[see][]{Cibirka2018}. As in previous works, we also mark for reference the corresponding areas $A(\mu > 5)$ and $A(\mu > 10)$ for the \textit{Hubble Frontier Field} (HFF) clusters \citep{Lotz2017}, computed from the \textsc{zitrin-ltm-gauss} models (in the full area provided for each cluster).
Even though RXJ0152 is a significantly smaller (less massive) lens, its lensing strength is nearly comparable to both the typical HFF or the RELICS clusters lensing strengths from both the LTM and Lenstool models. We also point that the LTM lensing strength is slightly greater than that from the Lenstool model, apart from the final modeling differences, because of the structure in the SW that we chose to not include in the Lenstool model due to the lack of lensing constraints in that region. This SW structure creates an additional region of high-magnification in our LTM model.
The high lensing efficiency of RXJ0152 is partly due to its merging state, where high-magnification regions arise between the merging subclumps or substructures projected on the plane of the sky \citep[i.e.][]{Torri2004,Meneghetti2007, Fedeli2010, Redlich2012, Wong2012, Cibirka2018}.\\
A primary goal of the RELICS observations was to detect a statistically significant number of high-redshift galaxies. \citet{Salmon2017} performed an extensive photometric study of the 41 RELICS cluster fields revealing 321 candidate galaxies with photometric redshifts between $z\sim6$ to $z\sim8$.

Particularly, the authors found large samples of such high-redshift candidate galaxies in fields magnified by relatively high-redshift (i.e. close to $z\sim 1$), morphologically complex clusters. Indeed, for a given lens angular-diameter distance $\mathrm{D_{L}}$, the lensing signal depends on the angular-diameter distances to the source ($\mathrm{D_{S}}$), and from the lens to the source ($\mathrm{D_{LS}}$), as $\propto \mathrm{D_{LS}}/$ $\mathrm{D_{S}}$. This means that for clusters at high redshift, the lensing power increases significantly with source redshift, relative to the slow increase in low redshift clusters. For instance, the highest redshift cluster ($z=0.972$) for which a full strong lens model was recently published, SPT-CLJ0615-5746 \citep{Paterno-Mahler2018}, based on RELICS observations. Their SL analysis reveals critical curves that substantially increase from $z_s=1.3$ to $z=9.93$ and together with RXJ0152 presented the second and third largest sample of high-redshift galaxies, respectively, within any RELICS cluster's field of view.
Similarly, \citep{Zitrin2013} carried out a lensing analysis with LTM on another famous galaxy cluster ACT-CLJ0102-49151 at a similarly high redshift, $z=0.87$, known as $\emph{El Gordo}$ (and also part of the RELICS sample). The authors found that the two central clumps, each forming its own modest critical curve for a source redshift $z_{s}=2$, rapidly increase with source redshift and the two critical regions merge together into a large elongated lens for sources at $z_{s}=9$. For such high redshift clusters, the power to lens $z\sim1-2 $ background galaxies is small, but increases rapidly for higher redshift sources. More recently, and adopting the fully parametric SL algorithm \textit{Lenstool} \citep{Jullo2007}, \citet{Cerny2017} found similar results where its Einstein radius increases from $\mathrm{R_E}=27.2\pm1.4"$ for a source at $z_s=3.0$ to $\mathrm{R_E}=40.3\pm2.0"$ at $z_s=9.0$ and leading to the discovery of the fourth largest sample of high-z galaxy candidates in RELICS.

The high-z candidates within RXJ0152's FOV are presented in Table ~\ref{table:highzcan} and their positions are indicated in the left panel of Figure ~\ref{mag}, mainly lying outside of the $z_s=6.5$ critical area. For each high-z candidate, we present a magnification estimate (and its statistical uncertainty) from our best-fit model. The absolute magnitude, $M_{uv}$, at $\lambda = 1500$ \AA, is then obtained following the UV continuum slope $f_{\lambda}\propto \lambda^\beta$ parametrization for galaxies \citep{Meurer1999} that we compute with a weighted least-squares fit using the four WFC3/IR bands (F105W, F125W, F140W, and F160W). The flux corresponding to the redshifted $\lambda = 1500$ \AA\ is then used to obtain the absolute magnitude, given by $M_{AB}=31.4-2.5\log_{10}(\mathrm{F_{nJy}})$.
As input we use the EAZY redshift \citep{Brammer2008} estimates given in \citet{Salmon2017}, which consistently predict these objects to be at high-z (the scope being to characterize the intrinsic properties of high-z candidates). The resulting rest-frame UV luminosities (corrected for lensing magnifications) have a mean of $\mathrm{M_{UV}}\sim-18.2(-19.0)$ and standard deviation of 1.03(0.75) for the samples at $z=6(7)$ respectively. We used our best-fit SL model to check (but did not find) for any high-redshift multiply-imaged galaxies. However, our SL model can provide hints about the true nature of the candidate CL0152-13-0505 which would more probably be a low-z galaxy since, at $z\sim5.6$, our model predicts further multiple images that we do not identify whereas for $z\sim1.0$, a solution predicted by the BPZ photometric code, the galaxy is not multiply-imaged.

\begin{table*}
	\caption{High-z ($z \sim 6-7$) lensed candidates}            
	\label{table:highzcan}      
	\centering  
    {\renewcommand{\arraystretch}{1.6}
	\begin{tabular}{c c c c c c c c c}        
		\hline\hline                 
		Galaxy ID\tablenotemark{a} & R.A. & Decl& $\mathrm{J_{125}}$ \tablenotemark{b}& $\mathrm{z_{phot}^{BPZ}}$\tablenotemark{c} & $\mathrm{z_{phot}^{EZ}}$\tablenotemark{d}& $\mu_{LTM}$ \tablenotemark{e}& $\mu_{Lenstool}$ \tablenotemark{f}& $M_{uv,1500}$\tablenotemark{g} \\  
		&[J2000]&[J2000]&[AB]&&&&&[AB]\\
		\hline          
CL0152-13-0152 & 28.1748725 & -13.9747007 & $ 27.21 \pm 0.16 $ &$5.9^{+0.3}_{-0.3} $ & $ 6.1^{+0.3}_{-0.4}$ & $1.45^{+0.03}_{-0.04}$ & $1.33\pm0.07$ &$-19.18^{+0.30}_{-0.31}$\\
CL0152-13-0207 & 28.1713376 & -13.9728768 & $ 27.37 \pm 0.16$& $1.0^{+5.3}_{-0.2} $ & $ 5.7^{+0.9}_{-4.5}$ &$1.88^{+0.11}_{-0.13}$&$1.53\pm0.09$&$-18.44^{+0.34}_{-0.79}$ \\
CL0152-13-0214 & 28.1733101 & -13.9726804 & $ 27.52 \pm 0.17 $& $5.5^{+0.2}_{-0.5} $ &$ 5.7^{+0.1}_{-0.7}$ &$1.74^{+0.11}_{-0.14}$& $1.49\pm0.09$ &$-18.09^{+0.33}_{-0.33}$  \\
CL0152-13-0391 & 28.1999315 & -13.9471136 & $ 27.15 \pm 0.20 $& $5.6^{+0.2}_{-0.3} $ &$ 5.8^{+0.1}_{-0.4}$ &$3.73^{+0.84}_{-0.65}$&$1.29\pm0.05$&$-18.12^{+0.30}_{-0.31}$  \\
CL0152-13-0505 & 28.1838405 & -13.9498063 & $ 27.70 \pm 0.19 $ &$1.1^{+5.2}_{-0.2} $ & $ 5.6^{+0.6}_{-4.6}$ &$5.49^{+1.60}_{-1.48}$& $4.09\pm1.25$ &$-16.37^{+0.32}_{-0.81}$  \\
CL0152-13-0608 & 28.1785995 & -13.950044 & $ 27.70 \pm 0.19 $ &$5.4^{+0.3}_{-0.3} $ & $ 5.7^{+0.2}_{-0.6}$ &$2.20^{+0.04}_{-0.04}$& $1.62\pm0.20$ &$-17.45^{+0.30}_{-0.32}$  \\
CL0152-13-0771 & 28.193064 & -13.9545314 &$ 25.62 \pm 0.06 $&$6.1^{+0.1}_{-0.2} $ &$ 6.0^{+0.2}_{-0.1}$&$2.18^{+0.08}_{-0.08}$&$1.77\pm0.35$&$-20.39^{+0.41}_{-0.37}$\\
CL0152-13-0800 & 28.1721392 & -13.9550721 & $ 26.99 \pm 0.13 $& $5.5^{+0.1}_{-0.3} $ & $ 5.6^{+0.2}_{-0.4}$ &$2.17^{+0.05}_{-0.05}$& $1.41\pm0.12$ &$-18.76^{+0.30}_{-0.31}$ \\
CL0152-13-0924 & 28.197796 & -13.957163 & $ 27.98 \pm 0.22 $& $5.4^{+0.2}_{-4.5} $ & $ 5.7^{+0.1}_{-1.7}$ &$2.03^{+0.20}_{-0.32}$& $1.17\pm0.05$&$-17.73^{+0.30}_{-0.41}$ \\
CL0152-13-1210 & 28.1864788 & -13.9628915 & $ 27.27 \pm 0.15 $& $6.1^{+0.4}_{-5.2} $ & $ 5.8^{+0.9}_{-4.8}$ &$1.82^{+0.06}_{-0.06}$& $1.41\pm0.12$ &$-18.44^{+0.33}_{-0.82}$ \\
CL0152-13-1307 & 28.1664822 & -13.9646117 & $ 27.28 \pm 0.19 $& $5.6^{+0.3}_{-4.9} $ & $ 5.7^{+0.4}_{-4.9}$ &$5.61^{+2.05}_{-1.76}$& $1.43\pm0.08$&$-17.13^{+0.31}_{-0.85}$ \\
CL0152-13-1341 & 28.1771307 & -13.9652886 & $ 28.15 \pm 0.25 $& $0.9^{+4.7}_{-0.4} $ & $ 5.6^{+0.3}_{-5.1}$ &$3.64^{+0.90}_{-0.83}$& $2.60\pm0.43$ &$-16.89^{+0.41}_{-0.93}$ \\
CL0152-13-1445& 28.179935 & -13.9671165 & $ 28.14 \pm 0.25 $& $5.5^{+0.2}_{-4.8} $ & $ 5.8^{+0.1}_{-5.0}$ &$1.77^{+0.10}_{-0.05}$& $1.59\pm0.16$ &$-17.97^{+0.35}_{-0.87}$ \\
CL0152-13-1494 & 28.1869999 & -13.9682532 & $ 27.53 \pm 0.18 $ &$5.6^{+0.3}_{-4.8} $ & $ 5.8^{+0.2}_{-5.1}$ &$1.41^{+0.03}_{-0.03}$& $1.21\pm0.06$ &$ -18.41^{+0.30}_{-0.87}$ \\
CL0152-13-1508 & 28.1656067 & -13.9686442 & $ 24.54 \pm 0.04 $ &$5.6^{+0.1}_{-0.1} $ & $ 5.5^{+0.3}_{-0.2}$ &$3.54^{+0.10}_{-0.09}$& $1.47\pm0.07$&$-20.35^{+0.31}_{-0.30}$ \\
CL0152-13-1546 & 28.1782025 & -13.9693313 & $ 28.04 \pm 0.21 $ &$5.4^{+0.4}_{-4.8} $ & $ 5.7^{+0.3}_{-4.8}$ &$1.67^{+0.04}_{-0.05}$& $1.51\pm0.12$&$-18.09^{+0.30}_{-0.83}$ \\
 CL0152-13-1569 & 28.1779564 & -13.9696918 & $ 28.15 \pm 0.25 $ &$0.9^{+4.6}_{-0.3} $ & $ 5.6^{+0.2}_{-4.6}$ &$1.65^{+0.04}_{-0.05}$&$1.49\pm0.12$ &$-18.19^{+0.30}_{-0.81}$ \\
CL0152-13-1576 & 28.1816193 & -13.9699027 & $ 27.15 \pm 0.14 $ &$5.5^{+0.2}_{-5.0} $ & $ 5.5^{+0.4}_{-4.9}$ &$1.47^{+0.03}_{-0.04}$& $1.31\pm0.08$ &$-18.79^{+0.31}_{-0.87}$ \\
CL0152-13-1642 & 28.1881258 & -13.9439934 & $ 27.95 \pm 0.22 $ &$5.8^{+0.4}_{-4.8} $ & $ 5.9^{+0.6}_{-4.9}$ &$5.14^{+1.30}_{-2.05}$&$1.94\pm0.17$&$-16.85^{+0.32}_{-0.83}$  \\
		\hline
CL0152-13-0191 & 28.1716411 &-13.9734429 & $ 27.13 \pm 0.28 $ &$ 6.6^{+0.5}_{-0.5} $ & $ 6.9^{+0.6}_{-0.7} $ &$1.78^{+0.06}_{-0.08}$ & $1.51\pm0.08$&$-19.10^{+0.31}_{-0.32}$ \\
CL0152-13-0259 & 28.1825175 & -13.9717095 & $ 26.36 \pm 0.23 $& $ 6.7^{+0.4}_{-0.4} $ & $ 7.0^{+0.5}_{-0.5} $ & $1.40^{+0.02}_{-0.03}$& $1.24\pm0.07$&$-20.16^{+0.31}_{-0.31}$ \\
CL0152-13-0410 & 28.1824016 & -13.9469211 & $ 26.58 \pm 0.24 $ &$ 6.4^{+0.2}_{-0.3} $ & $ 6.7^{+0.2}_{-0.4} $ & $2.58^{+0.04}_{-0.05}$& $1.75\pm0.24$&$-19.24^{+0.30}_{-0.31}$\\
CL0152-13-0525 & 28.1926776 & -13.9499311 & $ 27.19 \pm 0.29 $ &$ 6.8^{+0.7}_{-5.8} $ & $ 7.2^{+0.8}_{-5.9} $ &$4.02^{+0.11}_{-0.10}$ &$1.83\pm0.16$ &$-18.02^{+0.32}_{-0.84}$ \\
CL0152-13-1254 & 28.1806823 & -13.9635242 & $ 27.20 \pm 0.28 $ &$ 6.5^{+0.4}_{-5.5} $ & $ 6.9^{+0.3}_{-5.6} $ & $2.56^{+0.42}_{-0.33}$&$2.24\pm0.40$ &$-18.31^{+0.32}_{-0.83}$\\
		\hline\hline                                
	\end{tabular}}
      \tablecomments{}
      \tablenotetext{1}{Galaxy ID, following \citet{Salmon2017} notations. Note that the cluster is also named CL 0152.7-1357. The horizontal line separates candidates at $z\sim6$ and $z\sim7$.}
     \tablenotetext{2}{Apparent magnitude in the F125W band.}
  \tablenotetext{3}{Redshift estimation based on the BPZ pipeline along with their $1\sigma$ uncertainties.}
 \tablenotetext{4}{Redshift estimation based on the EAZY pipeline along with their $1\sigma$ uncertainties.}
  \tablenotetext{5}{Magnification estimates (at the respective source redshift) from our LTM best-fit model and the corresponding statistical uncertainty (measured as the standard deviation). The best-fit value is the one used for all relevant computations.}
    \tablenotetext{6}{Average magnification estimates (at the respective source redshift) from our Lenstool model and statistical uncertainty (i.e. the standard deviation) from 2000 MCMC models.}
  \tablenotetext{7}{Absolute magnitude, $M_{uv}$, at $\lambda = 1500$ \AA\ for which the errors have been propagated from the photometric and magnification uncertainties based on our best-fit LTM model.}
\end{table*}

Another factor that can enhance the lens efficiency of clusters is the high ellipticity or elongation of the lens. 
The effect of substructures and ellipticity was quantified, for example, in N-body simulated and realistic clusters in \citet{Meneghetti2007} where the authors found that substructures and cluster ellipticity account for $\sim30-40\%$ of the total cluster cross section, respectively. The combination of these factors explained for instance the unexpectedly high number density of multiple images seen in MACS J0416.1−2403 at $z = 0.40$ \citep[see][]{Zitrin2013a}. 
The mass distribution of RXJ0152 appears to be highly elongated for which we estimate an ellipticity (measured as $(a^2-b^2)/(a^2+b^2)$) of $\sim0.76\pm0.02$ in the inner regions but dropping to $\sim0.54\pm0.05$ at larger radius. This values are similar to those measured for MACS  J0416.1−2403 with an ellipticity of $\sim0.72\pm0.01(0.47\pm0.04)$ in the inner(outer) regions. These measurements are in agreement with results from N-body dark matter only simulations \citep{Despali2017}, where $10^{11}-10^{15} M_{\odot}h^{-1}$ haloes are more elongated in the centre than the outskirts, which still undergo significant interactions with the cluster's environmnent.
In MACS J0416.1−2403, \citet{Zitrin2013a} identified around 70 multiple images over a critical area of $\sim 0.6$ $arcmin^2$ (at $z=2$), from deeper imaging from the \textit{CLASH/HST} survey \citep[see][]{Postman2012}. 
The critical area for $z_s=2$ for RXJ0152 is only $\sim 0.1$ $arcmin^2$ where we find at least 25 multiple images in total (31, if also considering candidate identifications), so that the number of multiple images per area, i.e., the density of multiple images, is exceptionally high, enabling high-resolution constraints on its central mass distribution. 

The enhanced lensing efficiency of RXJ0152 also likely contributes in making the high-redshift, merging cluster RXJ0152 one of the most highly-magnifying lenses from the RELICS sample. That said, it should however also be noted that cosmic variance can play a non-negligible role in boosting the number of multiple images and high-redshift galaxies within a cluster's FOV \citep{Leung2018}. The uncertainty in the volume density of high-redshift galaxies arising from cosmic variance was estimated to be around $\sim10-20\%$ for Lyman-break galaxies at $z\sim3-4$ \citep{Somerville2004} but it can increase to $\sim35\%$  for higher-redshift sources \citep[around $z\sim5$,][]{Trenti2008}. Uncertainties associated with cosmic variance should carefully be taken into account in high-redshift studies, for example, as it can significantly affect the constraints on the faint-end slope of the high-redshift luminosity function \citep{Robertson2014}, which is beyond the scope of this study.

We also note that the RMS of our model can be artificially boosted. There are two main reasons for this. The first is technical: the LTM, given it is not fully analytic, is constructed on a grid. The grid's finite resolution (typically similar or of the order of the \textit{HST} pixel scale), due to round ups in the image and source positions, introduces a modest RMS boost which can reach 0.1-0.3 arcseconds per system (we now work on assessing this more thoroughly and more exact results will be reported in future work). The second reason is that the minimum of free parameters and the assumption that mass is coupled to the light distribution, while on one hand maximizing prediction power allowing for the detection of multiple images, does not allow for excessive flexibility in the model, and the fit is limited to the LTM assumption framework. Finally, we would like to emphasize that users should cautiously use SL modeling outputs (such as convergence, magnification, etc.) beyond the SL regime where multiple images are seen. The lens model is thus considered as an extrapolation beyond this limit. In addition, the smoothing and other interpolations used in our methodology can introduce artifacts at the edges of the modeled FOV.

Recent studies have focused on better understanding and quantifying the impact of systematic errors arising from different assumptions (i.e. different algorithms) in the modeling of strong lensing clusters \citep{Treu2016, Meneghetti2017,Bouwens2017}.
We have then modeled RXJ0152 with the Lenstool in order to compare the main SL outputs between the two modeling tools which are the main algorithms providing SL models of RELICS clusters for the community. 

We find that the resulting mass distributions are in good agreement as shown in Figure ~\ref{mass}. While the Lenstool model estimates a higher mass in the inner region of the cluster core (the LTM being shallower as is typically the case) and the LTM model is more massive in the outskirts due to the structure in the SW, both models are in very good agreement in the intermediate region. 

Regarding the high-z candidates, the magnification estimates are in fairly good agreement between the two models but the discrepancies between models become larger for high magnification values \citep{Bouwens2017}. The LTM pipeline predicts some high-redshift candidates with significantly higher magnification values. This is partly ecause of some high-magnification region between the merging clump- that we did not include in the Lenstool model.

\section{Summary}\label{sec:summary}
The merging galaxy cluster RXJ0152.7-1357 (also known as CL 0152.7-1357; $z=0.83$) is one of the X-ray brightest, and thus best studied clusters, at such high-redshifts, but missing a full, public SL analysis to date.

In this work we have presented a SL analysis of RXJ0152 based on recent observations from the RELICS survey and adopting a Light-Traces-Mass methodology that allowed us to uncover several sets of multiple images of background galaxies to be used as constraints for the modeling. 
\citet{Umetsu2005} had uncovered, and spectroscopically confirmed one multiply-imaged system that allowed them to put constraints on the NE clump of the cluster. Thanks to the RELICS survey we were now able to uncover a relatively large number of new multiple images over a small area of just $\sim 0.4$ $arcmin^2$, allowing us to put high-resolution constraints on the central mass distribution of RXJ0152.7-1357.

The mass distribution of RXJ0152's core, as revealed by our SL modeling and as indicated by the member galaxy distribution, shows a clumpy morphology made of several substructures which further supports the merging scenario reported in previous works \citep{Maughan2003, Jee2005}.
RXJ0152 appears to be a modest lens with relatively small critical curves for a source redshift $z\sim2$, over several merging clumps, and enclosing a mass of $2.5\pm0.4 \times 10^{13} M_{\odot}$. We note that, together with the \textit{Baby Bullet} cluster, RXJ0152 was one of the 2 RELICS clusters not detected in the SZ mass Planck cluster catalog PSZ2 \citep{Planck2015}. For higher redshift sources ($z\sim9$) these critical curves merge, boosting the critical area. As a result, RXJ0152.7-1357 presents a sightly smaller, but overall similar lensing strength for $z\sim9$ sources, as other RELICS cluster that were found to be massive and prominent lenses (see Figure ~\ref{mag}).

The elongated substructure chain composing RXJ0152 results in an efficient lens \citep[e.g.][]{Merten2011,Limousin2012}, accounting in part for the high density of multiply imaged galaxies. RXJ0152 also comprises the third-larger sample of high-redshift ($z\sim6-7$) candidates among all clusters in the RELICS program  \citep{Salmon2017}, a sample which we further characterized in this work thanks to our lens model.
This RELICS cluster shows the advantage of targeting high-redshift, merging clusters, even if modestly massive, as the high-level of substructure together with its elongated shape boosts the lensing efficiency \citep{Zitrin2013}.

Finally, we present a comparison between the LTM and Lenstool SL models. In general, both the mass distribution and magnification values are in good agreement, the differences arising from the distinct assumptions in the modeling techniques and parametrization. The LTM modeling estimates a larger high-magnification area due to the SW structure. 
A more detailed comparison between these two techniques is relegated to a future paper. 

Among all massive galaxy clusters observed with \textit{HST} to date, there is a significant number of clusters having none or few spectroscopically measured multiple images. Similarly, RXJ0152 has only one spectroscopically measured system, presented in \citet{Umetsu2005}. Upcoming observing campaigns will help overcome this source of systematic uncertainties, probably the main caveat of current SL models \citep{Johnson2016, Remolina2018}.\\
The lens models presented in this work, as well as magnification maps, are made publicly available through the MAST archive\textsuperscript{\ref{mast}}.\\

\acknowledgments
We kindly thank the anonymous referee for her/his suggestions that helped to improve the paper.
A.A thanks Jean-Paul Kneib and Eric Jullo for useful discussions. This work is based on observations taken by the RELICS Treasury Program (GO-14096) with the NASA/ESA \textit{HST}. Program GO-14096 is supported by NASA through a grant from the Space Telescope Science Institute, which is operated by the Association of Universities for Research in Astronomy, Inc., under NASA contract NAS5-26555. This work was performed in part under the auspices of the U.S. Department of Energy by Lawrence Livermore National Laboratory under Contract DE-AC52-07NA27344. This paper includes data gathered with the 6.5 meter Magellan Telescopes located at Las Campanas Observatory, Chile. We thank Ian Roederer for executing the Magellan observation.  
K.U. acknowledges support from the Ministry of Science and Technology of Taiwan under the grant MOST 106-2628-M-001-003-MY3. RCL acknowledges support from an Australian Research Council Discovery Early Career Researcher Award (DE180101240). This work is partially supported by the Australian Research Council Centre of Excellence for All-Sky Astrophysics in 3 Dimensions (ASTRO-3D). S.T. acknowledges support from the ERC Consolidator Grant funding scheme (project ConTExt, grant No. 648179). The Cosmic Dawn Center is funded by the Danish National Research Foundation.

%

\vspace{5mm}



\bibliographystyle{apj}
\bibliography{test}
\appendix

\newpage
\begin{appendices}
 \section{Reproduction of the multiple images}
 
 \begin{figure*}[h!]
	\centering
	\includegraphics[width=0.85\linewidth]{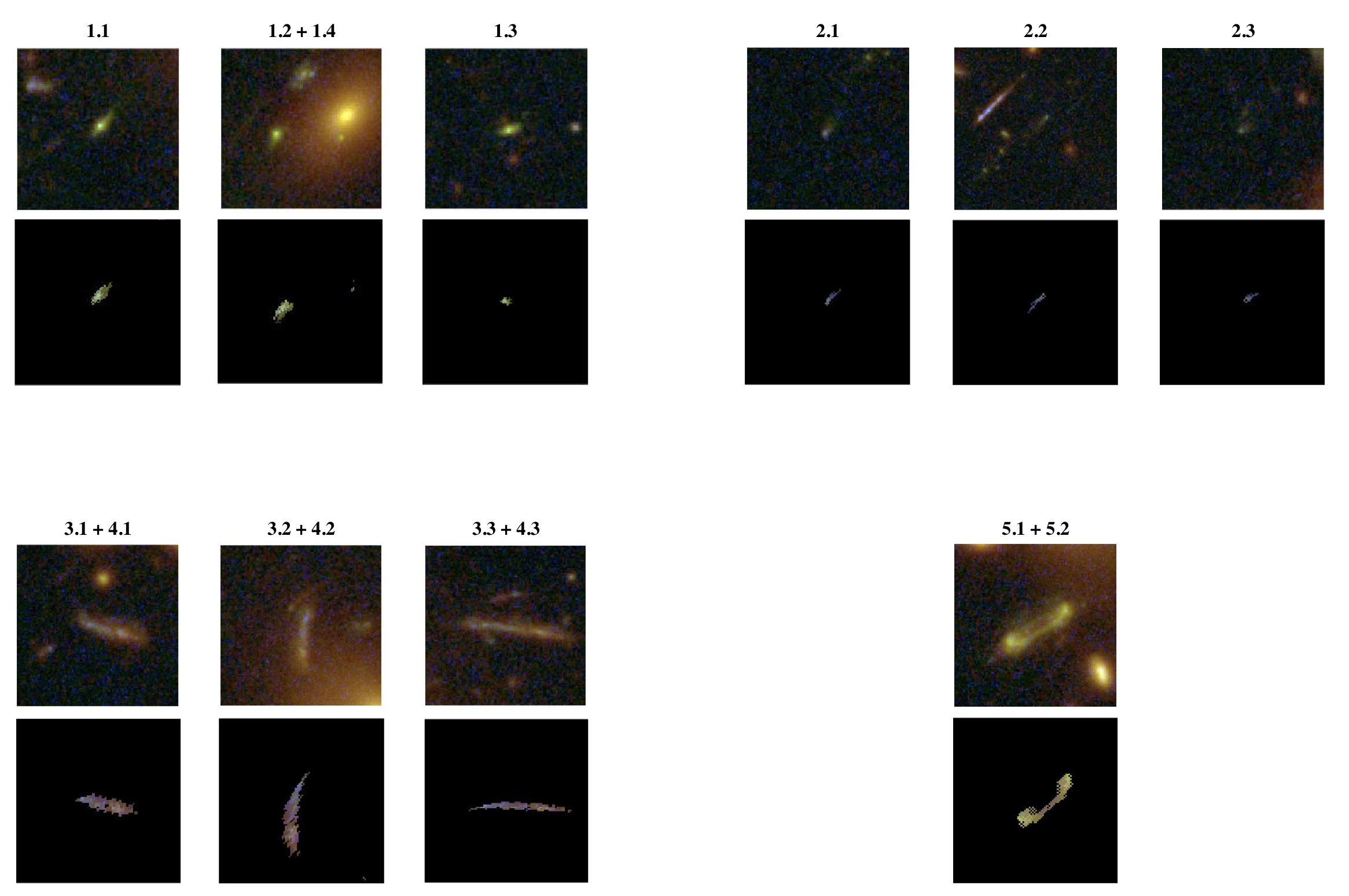} \\
   	\includegraphics[width=0.85\linewidth]{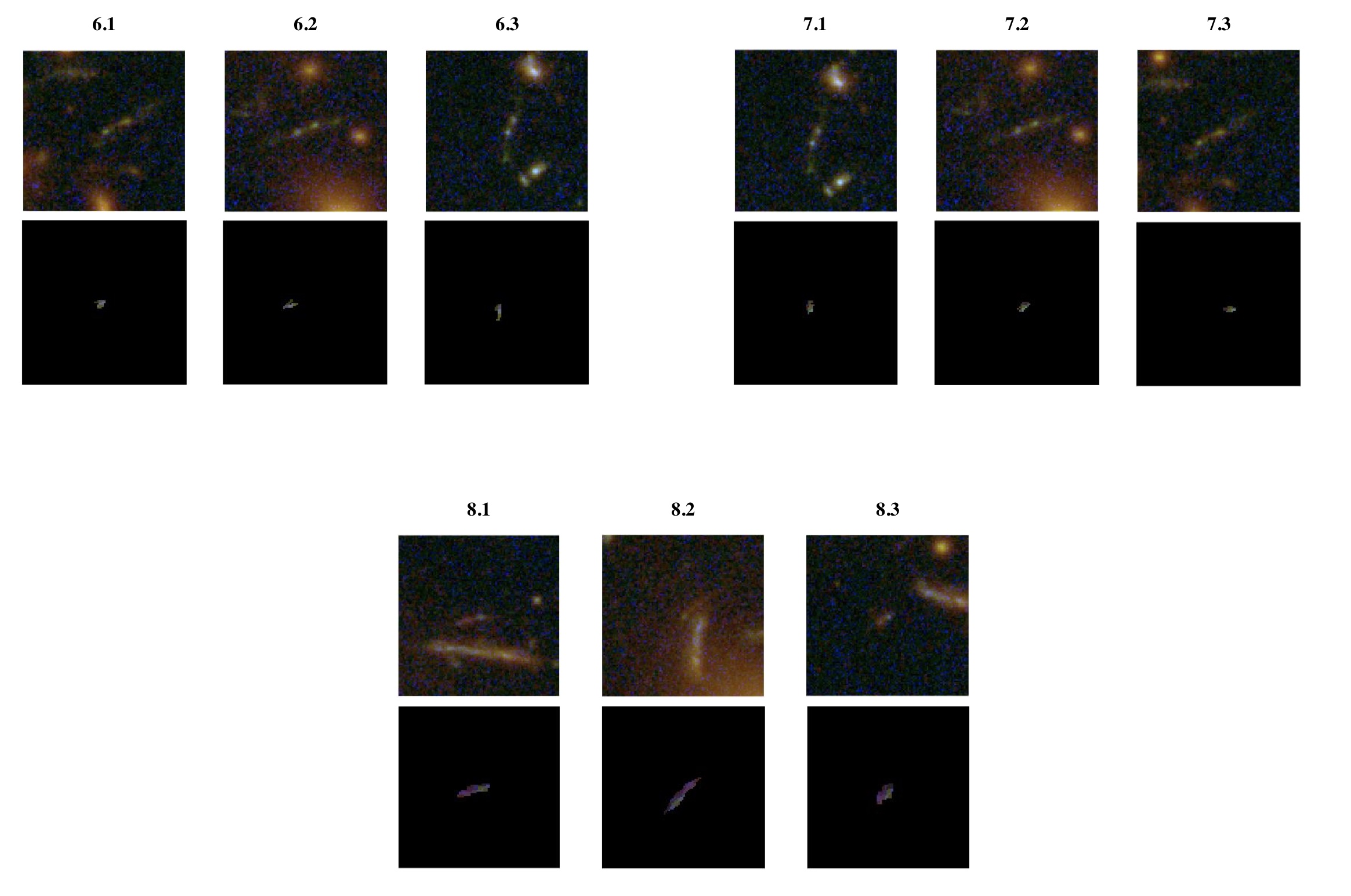}
	\caption{Reproduction of multiple images by our best-fit LTM model for RXJ0152. For each image, we de-lens the first image of the system to the source plane and back to the image plane to be compared to the other images of that system. The orientation and internal details of the model-predicted images (bottom rows) are similar to those of the observed images (upper rows).}
	\label{stamps0308}
\end{figure*}

\newpage

\section{RXJ0152's Lenstool model}\label{lenstool}
 
\begin{figure*}[h!]
	\centering
    \includegraphics[width=0.65\linewidth]{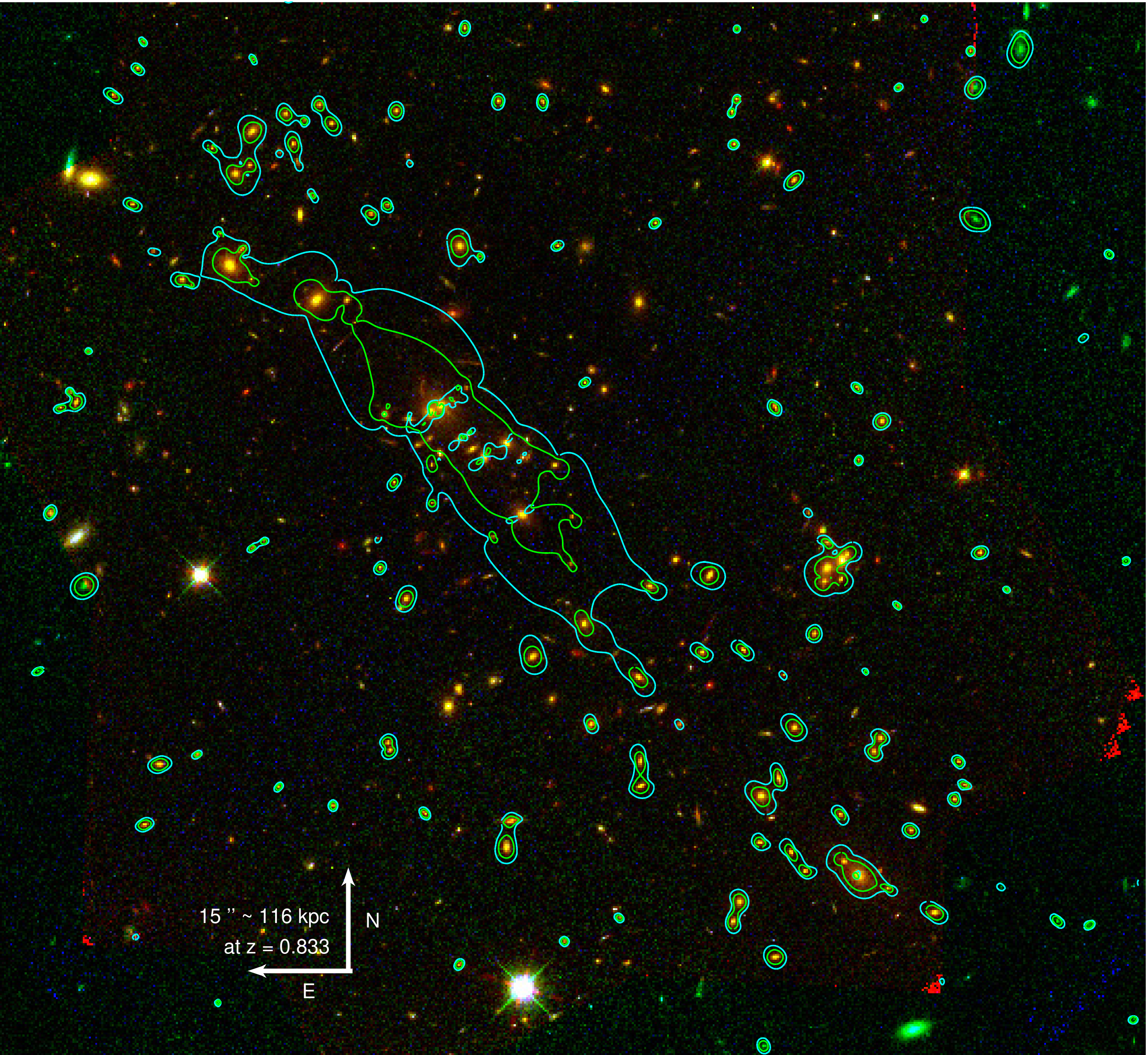}
	\caption{Critical curves from our best-fit Lenstool model at redshifts $z_s=2$ and $z_s=9$ in green and cyan, respectively.}
	\label{cc_LT}
\end{figure*}

\begin{figure*}[h!]
	\centering
	\includegraphics[width=0.6\linewidth]{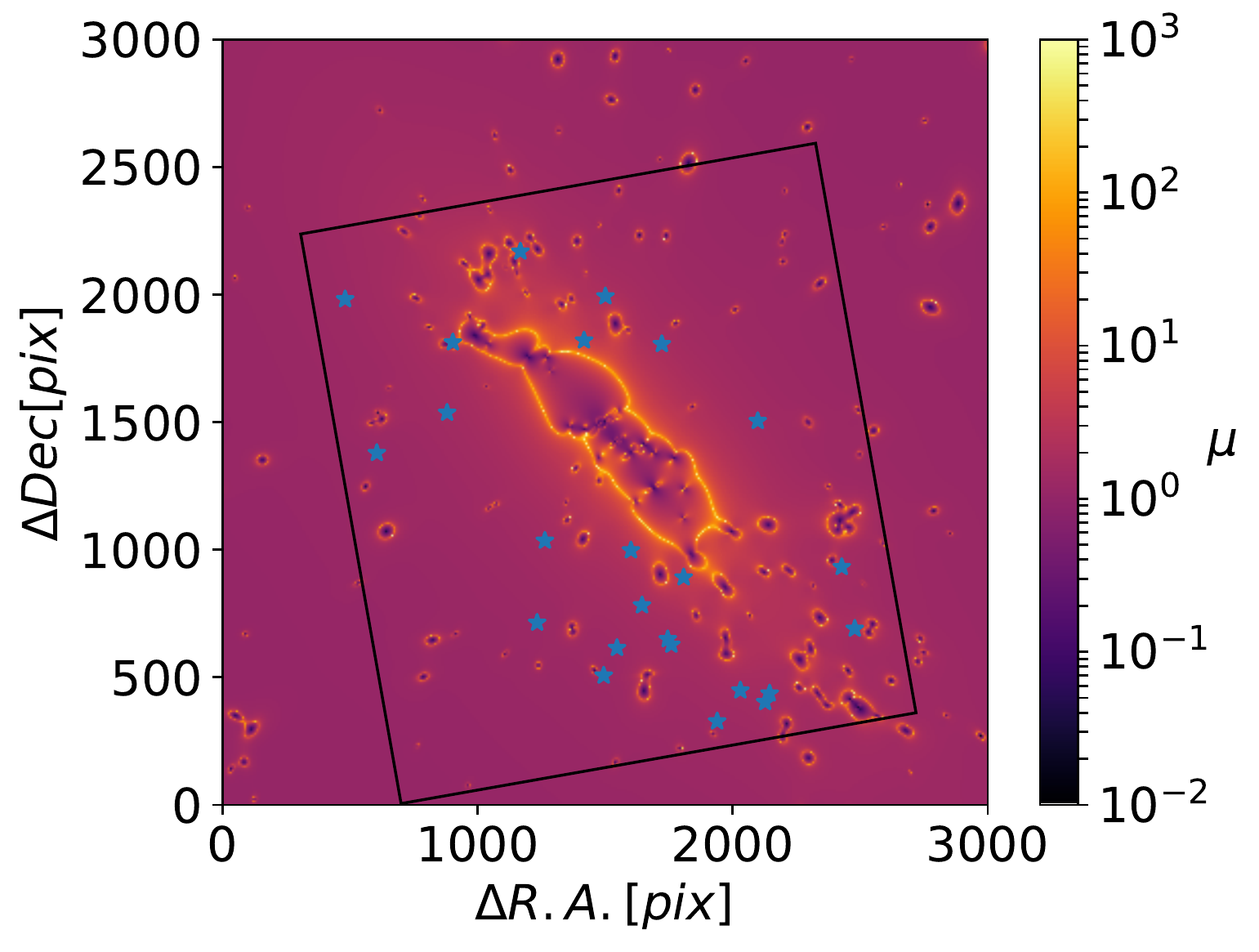}
	\caption{Left panel: Magnification map from our best-fit Lenstool model for a source at $z_s=6.5$. Same symbols as in Figure ~\ref{mag}.}
	\label{mag_LT}
\end{figure*} 

\newpage

\section{Best-fit parameters}\label{BFTab}
 
\begin{table*}[h!]
	\caption{Best-fit parameters from the LTM model.}            
	\label{table:BFltm}      
	\centering  
        {\renewcommand{\arraystretch}{1.4}
	\begin{tabular}{c | c c c c c c }        
		\hline\hline                 
		Component & q\tablenotemark{a} & s\tablenotemark{a}& $k_{new}$\tablenotemark{a} & $k_{gal}$\tablenotemark{a} & $\gamma$\tablenotemark{b} & $\phi$\tablenotemark{b} \\  
	  \hline \\         
		 Total Mass Distribution&  $1.40^{+0.03}_{-0.04}$ & $88.0^{+7.0}_{-6.0}$ & $1.21^{+0.09}_{-0.10}$ & $0.15^{+0.04}_{-0.03}$ & -&-\\ \\
    \hline  \\
    	External Shear & - & - &-&-&$0.05^{+0.4}_{-0.04}$ &$0.50^{+0.35}_{-0.37}$\\ \\
		\hline\hline                                
	\end{tabular}}
         \tablecomments{}
      \tablenotetext{1}{Best fit values with 1$\sigma$ uncertainties of the basic LTM parameters predsented in Section \ref{sec:lens_modelLTM}.}
     \tablenotetext{2}{Amplitude and position angle of the external shear with 1$\sigma$ uncertainties.}
\end{table*}

\begin{table*}[h!]\label{table:BF2}
\centering
\caption{Best-fit parameters from the Lenstool model.}
\label{tab:fit_param}
{\renewcommand{\arraystretch}{2.4}
\begin{tabular}{r|cccccccc}
\hline\hline 
 Component & $\Delta\alpha^{\rm ~a}$ & $\Delta\delta^{\rm ~a}$ & $\varepsilon^{\rm ~b}$ & $\theta^{\rm ~c}$ & $\sigma_0^{\rm ~d}$ & r$_{\rm cut}^{\rm ~d}$ & r$_{\rm core}^{\rm ~d}$\\ 
  & [\arcsec] & [\arcsec] &   & [$\deg$] & [km\ s$^{-1}$] & [kpc] & [kpc]\\ 
\hline 
 DM & $3.7^{+2.0}_{-1.1}$ & $-4.7^{+1.5}_{-2.0}$ & $0.74^{+0.04}_{-0.08}$ & $130.0^{+1.2}_{-1.2}$ & $926.1^{+41.1}_{-40.8}$ & $[1500.0]^{\rm ~e}$ & $8.7^{+2.5}_{-1.7}$\\ 
 $L^{*}$ Galaxy & -- & -- & -- & -- & $276.5^{+18.0}_{-19.4}$ & -- & --\\ 
\hline\hline
\end{tabular}}
      \tablecomments{}
      \tablenotetext{1}{Positional offsets with respect to the reference point (R.A=28.183021 deg; Decl=-13.955764 deg).}
     \tablenotetext{2}{Ellipticity.}
     \tablenotetext{3}{Position angle.}
      \tablenotetext{4}{PIEMD best-fit parameters.}
      \tablenotetext{5}{Fixed value.}
\end{table*}

\end{appendices}


\end{document}